\documentclass{PoS}

\usepackage{amsmath,amssymb,float}

\newcommand{\be}{\begin{equation}}
\newcommand{\ee}{\end{equation}}
\newcommand{\ba}{\begin{array}}
\newcommand{\bqa}{\begin{eqnarray}}
\newcommand{\eqa}{\end{eqnarray}}

\title{Recent developments in $U(3)$ $\chi$PT: meson-meson scattering and finite energy sum rules}

\ShortTitle{Recent developments in $U(3)$ $\chi$PT: Scattering and FESR }

\author{\speaker{Zhi-Hui~Guo} \\ 
        Department of Physics, Hebei Normal University, 050024 Shijiazhuang, P.~R.~China \\ 
    and Departamento de F\'isica, Universidad de Murcia,  E-30071 Murcia, Spain. \\ 
        E-mail: \email{guo@um.es}}

\author{J.A.~Oller\\
       Departamento de F\'isica, Universidad de Murcia,  E-30071 Murcia, Spain.\\
        E-mail: \email{oller@um.es}}

\author{J.~Ruiz de Elvira\\
       Departamento de F\'isica Te\'orica II, Universidad Complutense de Madrid, E-28040 Madrid, Spain.\\
        E-mail: \email{jacobore@rect.ucm.es}}

\abstract{ We discuss the meson-meson scattering and finite energy sum rule (FESR), based on the 
one-loop calculation within $U(3)$ chiral perturbation theory ($\chi$PT).  
First we obtain the pertinent resonance spectroscopy from the unitarized partial wave scattering amplitudes. Then we investigate how 
well the FESR can be satisfied in the physical situation at $N_C=3$. Further discussions on the extrapolation 
of $N_C$ are also given. 
 }

\FullConference{Sixth International Conference on Quarks and Nuclear Physics,\\
		April 16-20, 2012\\
		Ecole Polytechnique, Palaiseau, Paris}

\begin{document}

\section{Introduction}

Scalar dynamics of the light quark flavors in the nonperturbative energy region (especially below 1 GeV) 
has been quite a topical research subject in the last decade, see Refs.~
\cite{guo11prd,guo12plb,scalarref,oller99prd}. Despite the confirmation of the mass and width of the broad resonances 
by different groups~\cite{pdg}, their nature is still under a vivid debate. 
In the recent works~\cite{guo11prd,guo12plb}, we make a comprehensive study of the scalar resonances below 1.5 GeV. 
Here we mainly discuss the results on the meson-meson scattering 
and FESR (a way to quantify the semi-local or average duality)~\cite{collinsbook}.

First, the meson-meson scattering amplitudes are unitarized to fit to experimental data and the relevant resonance spectroscopy 
is confirmed. After then, we further investigate the role of resonances in semi-local duality, 
not only for the physical case at $N_C=3$, but also for the situation with $N_C>3$. 
The fulfillment of the nontrivial aspects of finite energy sum rules poses a strong support of the proposed picture for 
the scalar dynamics in our work.

\section{Theoretical framework}

Our theoretical framework is $U(3)$ $\chi$PT~\cite{herrera97npb,kaiser00epjc}, which includes not 
only the pseudo-Goldstone octet $\pi$, $K$ and $\eta_8$, as $SU(3)$ $\chi$PT~\cite{gasser85}, but also the singlet $\eta_1$. 
After the diagonalization of the $\eta_8$-$\eta_1$ mixing, one has the physical $\eta$ and $\eta'$ mesons as the dynamical degrees of 
freedom in  $U(3)$ $\chi$PT. This qualifies $U(3)$ $\chi$PT as an improved approach for studying the $\eta$ and $\eta'$ dynamics 
compared to the $SU(3)$ version. 
Instead of considering the contributions from the higher order low energy constants (LECs), we introduce explicitly 
the resonance exchanges at tree level, assuming the resonance saturation of LECs in our work. 
We follow the resonance chiral theory~\cite{ecker89npb} to include the scalar, vector and pseudoscalar resonance contributions. 
In addition we take into account another two local chiral operators $\delta L_8$ 
and $\Lambda_2$ in the calculation~\cite{guo11prd,guo12plb}.

We calculate the mass and wave function renormalizations, pion decay constant and all of the meson-meson scattering processes 
in $U(3)$ $\chi$PT up to one-loop order plus the tree level exchange of resonances. 
The perturbative amplitudes with definite isospin $I$ and angular momentum $J$ are then unitarized through the nonperturbative N/D approach~\cite{oller99prd} 
\begin{align}\label{defunitarizedT}
T^{IJ}(s)&=\big[ 1 + N^{IJ}(s) \, g^{IJ}(s) \big]^{-1} N^{IJ}(s)\,,
\end{align} 
where $g^{IJ}(s)$ takes care of the nonperturbative resummation of the unitarity cuts and all the 
contributions from the crossed-channel cuts are collected in $N^{IJ}(s)$.  
The explicit forms for $g^{IJ}(s)$ and $N^{IJ}(s)$ are given in Ref.~\cite{guo11prd}. 
With the unitarized partial waves in Eq.~\eqref{defunitarizedT}, it is straightforward to calculate the 
phase shift, inelasticity and invariant mass distribution, which can be used to fit experimental data.  

In addition to the phenomenological aspects, there also exist theoretical constraints on the scattering amplitudes. 
We focus on the semi-local duality in $\pi\pi$ scattering between the Regge theory and  hadronic degrees of freedom in this work. 
One of the objects to quantify the semi-local (or average) duality is the FESR 
\begin{equation}\label{defseml}
 \int_{\nu_1}^{\nu_2} \nu^{-n}\, {\rm Im}\, T_{\rm t, Regge }^{(I)}(\nu, t) \, d\nu =  
\int_{\nu_1}^{\nu_2} \nu^{-n}\, {\rm Im}\, T_{\rm t, Hadrons}^{(I)}(\nu, t) \, d\nu \,,
\end{equation}
with $\nu = \frac{s-u}{2}= \frac{2s + t-4m_\pi^2}{2}$ and $s,t,u$ the standard Mandelstam variables. 
In $\pi\pi$ scattering, the relations between the $t$- and $s$-channel amplitudes  with definite isospin $I$ are~\cite{collinsbook}
\begin{align}\label{tsrelation}
T_{\rm t}^{(0)}(s,t) &= \frac{1}{3} T_{\rm s}^{(0)}(s,t) + T_{\rm s}^{(1)}(s,t) + \frac{5}{3} T_{\rm s}^{(2)}(s,t) \,, \nonumber \\
T_{\rm t}^{(1)}(s,t) &= \frac{1}{3} T_{\rm s}^{(0)}(s,t) + \frac{1}{2} T_{\rm s}^{(1)}(s,t) - \frac{5}{6} T_{\rm s}^{(2)}(s,t) \,, \nonumber \\
T_{\rm t}^{(2)}(s,t) &= \frac{1}{3} T_{\rm s}^{(0)}(s,t) - \frac{1}{2} T_{\rm s}^{(1)}(s,t) + \frac{1}{6} T_{\rm s}^{(2)}(s,t) \,, 
\end{align}
where the subscript of $T$ labels the $t$- or $s$-channel and the superscript stands for the isospin $I$.   
The left hand side of Eq.~\eqref{defseml} can be evaluated in Regge theory. 
The explicit expressions and results can be found in Ref.~\cite{pelaez11prd} and references therein. 
For the right hand side of Eq.~\eqref{defseml}, we decompose the isospin amplitudes into a sum of partial wave amplitudes 
\begin{equation}\label{pwdecompose}
  {\rm Im}\, T_{\rm s}^{(I)}(\nu, t) = \sum_J (2 J + 1)\,  {\rm Im}\, T^{IJ}(s) \, P_J(z_s)\,,
\end{equation}
with $z_s = 1 + 2 t /(s - 4m_\pi^2)$ the cosine of the scattering angle in the $s$-channel 
center of mass frame and $P_J(z_s)$ the Legendre polynomials. 
By combining  Eqs.~\eqref{pwdecompose} and \eqref{tsrelation}, 
we can first obtain ${\rm Im}\, T_{t,{\rm Hadrons} }^{(I)}(\nu, t)$ and then compare 
the hadronic contributions with those from Regge theory.

One expects that the ``averaging'' in Eq.~\eqref{defseml} should start to work  at least for one resonance tower and this means the 
integration range $\nu_2 - \nu_1$ should be set as a multiple of 1~GeV$^2$. In this work we are interested in the energy region below 2~GeV$^2$. 
To proceed the discussion, it is advisory to consider the following ratios to cancel the uncertainties caused by the Regge couplings 
~\cite{guo12plb,pelaez11prd} 
\begin{eqnarray}\label{defRratio}
 R^I_n &=& \frac{ \int_{\nu_1}^{\nu_2} \nu^{-n}\, {\rm Im}\, T_{\rm t}^{(I)}(\nu, t)\, d\nu}
{\int_{\nu_1}^{\nu_3} \nu^{-n}\, {\rm Im}\, T_{\rm t}^{(I)}(\nu, t)\, d\nu}\,,  
 \\ \label{defFratio}
 F_n^{I I'} &=& \frac{ \int_{\nu_1}^{\nu_{\rm max} } \nu^{-n}\, {\rm Im}\, T_{\rm t}^{(I)}(\nu, t)\, d\nu}
{\int_{\nu_1}^{\nu_{\rm max}} \nu^{-n}\, {\rm Im}\, T_{\rm t}^{(I')}(\nu, t)\, d\nu}\,,
\end{eqnarray}
with $\nu_1$ the $\pi\pi$ threshold, $\nu_2=1$ GeV$^2$, $\nu_3=2$ GeV$^2$ and ${\nu_{\rm max}}=2$ GeV$^2$ in later discussions.

\section{Discussions}

We fit the unknown parameters in our theory to a large amount of experimental data, 
consisting of phase shifts and inelasticities of $\pi\pi \to \pi\pi (K\bar{K})$ and  
$\pi K \to \pi K$ scattering, with different isospin  and angular momentum numbers, and also the invariant mass distribution of 
the $\pi\eta$ system ~\cite{guo11prd,guo12plb}. The fit quality is fairly good and with the fitted parameters we obtain 
seven scalar and three vector resonances from our unitarized scattering amplitudes in the complex energy plane: 
$f_0(600)$, $f_0(980)$, $f_0(1370)$, $K^*_0(800)$, $K^*_0(1430)$, $a_0(980)$, $a_0(1450)$, $\rho(770)$, $K^*(892)$ and $\phi(1020)$. 
Their masses and widths agree well with the PDG values. In addition, we also calculate the coupling strengths of the 
resonances to the pseudo-Goldstone boson pairs. This comprises the couplings of the $f_0$ resonances to $\pi\pi$, 
$K\bar{K}$, $\eta\eta$, $\eta\eta'$ and $\eta'\eta'$, the couplings of $K^*_0$ to $K\pi$, $K\eta$ and $K\eta'$, the couplings 
of $a_0$ resonances to $\pi\eta$, $K\bar{K}$ and $\pi\eta'$ and also the relevant coupling strengths for the vector resonances. 
One important lesson we learn is that the $f_0(600)$ 
resonance is marginally coupled to $\eta$ and $\eta'$ mesons, indicating its insensitivity to the $\eta$ and $\eta'$ dynamics. 
The explicit numbers can be found in Refs.~\cite{guo11prd,guo12plb}. 

\begin{table}[ht]
\renewcommand{\tabcolsep}{0.05cm}
\renewcommand{\arraystretch}{1.2}
\begin{center}
{\small 
\begin{tabular}{|l|l|llll|llll|llll|llll|}
\hline
 & n &&  $R_n^0$   && $R_n^0$  && $R_n^1$  && $R_n^1$ && $F_n^{21}$  && $F_n^{21}$  
\\
 &   &&  $t=t_{\rm th}$   && $t=0$   && $t=t_{\rm th}$  && $t=0$  &&  $t=t_{\rm th}$  && $t=0$
\\
[0.1cm] 
\hline
\hline
Regge & 0 && 0.225 && 0.233 &&  0.325 &&  0.353  &&  $\sim 0$  &&  $\sim 0$   
\\[0.1cm] 
      & 1 && 0.425 && 0.452 &&  0.578 &&  0.642  &&   $\sim 0$ &&  $\sim 0$  
\\[0.1cm] 
      & 2 && 0.705 && 0.765 &&  0.839 &&  0.908  &&   $\sim 0$ &&  $\sim 0$   
\\[0.1cm] 
      & 3 && 0.916 && 0.958 &&  0.966 &&  0.990  &&   $\sim 0$ &&  $\sim 0$  
\\[0.1cm]   
\hline 
Ours  & 0   && 0.669 && 0.628 &&  0.836 &&  0.817 &&  -0.113 &&  0.040    
\\[0.1cm] 
$S+P$   & 1 && 0.837 && 0.812 &&  0.919 &&  0.908 && -0.230   &&  -0.087   
\\[0.1cm] 
Waves & 2   && 0.934 && 0.924 &&  0.966 &&  0.962 && -0.129   &&  0.028   
\\[0.1cm]  
      & 3   && 0.979 && 0.976 &&  0.989 &&  0.988 &&   0.169 &&   0.345  
\\[0.1cm]   
\hline 
Ours  & 0   && 0.410  && 0.400  &&  0.453  &&  0.468  &&  0.531  &&   0.587    
\\[0.1cm] 
$S+P+D$ & 1 && 0.653  && 0.643  &&  0.694  &&  0.706  &&  0.154  &&   0.236   
\\[0.1cm] 
Waves & 2   && 0.850  && 0.844  &&  0.875  &&  0.882  &&  0.027  &&   0.155    
\\[0.1cm] 
      & 3   && 0.954  && 0.953  &&  0.965  &&  0.968  &&  0.225  &&   0.388  
\\[0.1cm]  
\hline 
\hline
\end{tabular}
}
 \caption{ {\small Values for $R_n^I$ and $F_n^{II'}$. Results by using two values of $t$, 0 and $t_{{\rm th}}=4m_\pi^2$, are shown in the table. }
 }  
\label{tab:ratiosatnc3}
\end{center}
\end{table}

Now, it is interesting to explore the scattering amplitudes, which contain the relevant resonances, in finite energy sum rules.  
We compare the results from Regge theory and the unitarized $U(3)$ $\chi$PT amplitudes in Table~\ref{tab:ratiosatnc3}. 
When considering the $D$-wave contributions, we include the tensor resonances in meson-meson scattering following the framework 
in Ref.~\cite{ecker07epjc}. The first lesson we can learn from the numbers in Table \ref{tab:ratiosatnc3} is that 
the results are stable when switching from $t=0$ to $t=4m_\pi^2$. 
Next, let us focus on the ratios $R^{I_{\rm t}}_n$ with $I_{\rm t}=0,1$ and our current results qualitatively 
agree with the conclusions in Ref.~\cite{pelaez11prd}. 
The semi-local duality with $n=3$ can be well satisfied by just $S$- and $P$-waves, while the fulfillment 
for the $n=2$ case is marginal. About the cases with smaller values of $n$, higher partial 
waves and cut-offs are needed in order to fulfill the semi-local duality. As one can see in Table \ref{tab:ratiosatnc3}, the inclusion 
of the $D$-waves clearly improves the situations for $n=0,1$. 

Concerning the ratios $F^{21}_n$ with the isotensor amplitude involved, we observe different results. Due to the suppression of 
the Regge exchanges in $I_{\rm t}=2$ amplitude, the Regge theory predicts almost vanishing values for $F^{21}_n$. 
From Eq.~\eqref{tsrelation}, one can simply conclude that if the $P$-($S$-) wave contributions are dropped  
the ratios of $F^{21}_n$ should approach to +1(-1), which can be viewed as the criteria to claim the extreme violation of semi-local duality.  
By only considering the $S$- and $P$-waves, we find that semi-local duality is well satisfied for all values of $n$ and 
there is no clear sign that it is better satisfied for a specific value of $n$, contrary to the situations of $R^{I}_n$. 
While the introduction of the $D$-waves,  
instead of improving the situation, deteriorates the fulfillment of the semi-local duality for $n=0$.

\begin{figure}[ht]
\begin{center}
 \includegraphics[angle=0, width=0.80\textwidth]{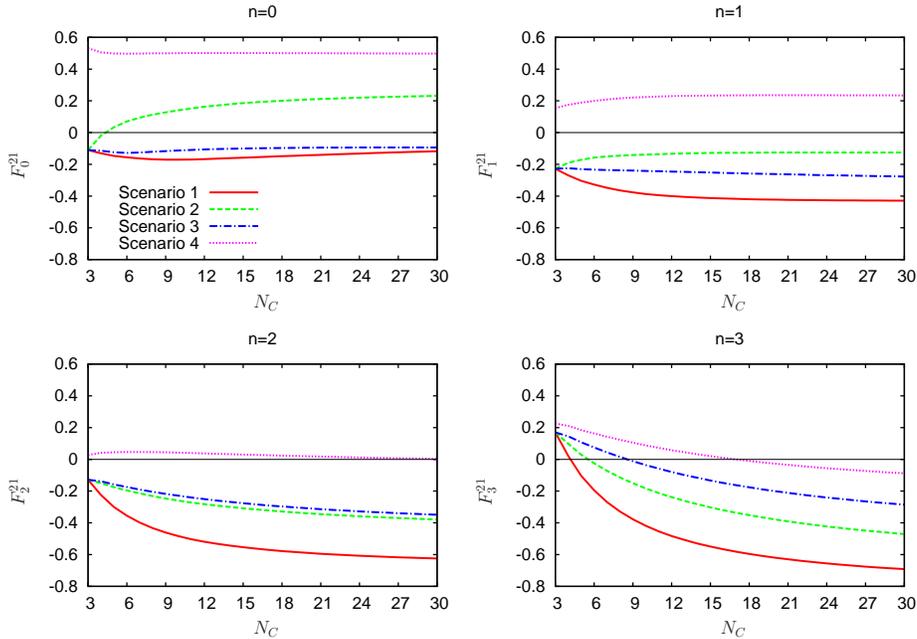}
\caption{{\small $F_n^{21}(t=4m_\pi^2)$. Detailed definitions for the different scenarios can be found in Ref.~\cite{guo12plb}. }}
\label{fig.ncdualityf21}
\end{center}
\end{figure}

All the discussions above are made in the physical situation, i.e. at $N_C=3$. It is interesting to investigate 
the fulfillments of semi-local duality by extrapolating the values of $N_C$ to larger numbers~\cite{pelaez11prd}.  
In order to perform this study, we need to know the $N_C$ scaling of the parameters in our theory. 
For the $N_C$ scaling at leading order, it is known without any ambiguity, which can be found in Refs.~\cite{guo11prd,ecker89npb}. 
While if one wants to consider the sub-leading order $N_C$ scaling, extra assumptions need to be provided. 
In order to estimate the uncertainties of the sub-leading order effects, 
we propose four different scenarios to discuss semi-local duality. It is interesting to point out that 
semi-local duality indeed can distinguish different scenarios and hence provides us a way to constrain the 
$N_C$ evolution of parameters~\cite{guo12plb}. This also gives us a valuable onset to study the $N_C$ evolution of 
the resonance poles and spectral sum rules~\cite{guo12plb}. 
In Fig.~\ref{fig.ncdualityf21}, we show how the different scenarios can be distinguished in the study of the ratios $F^{21}_n$. 
Taking into account the criteria we mentioned above, it is easy to conclude that Scenario 3 is the best one to satisfy semi-local 
duality. In Scenario 4, we take into account the $D$-wave contributions, in addition to the $S$- and $P$-waves that are included 
in Scenarios 1-3. The reason why Scenario 4 is disfavoured is due to the fact that the $D$-wave overbalances the vector resonance contributions
and hence it leads to too large values for the $n=0$ case. 
In the physical situation, as we mentioned before, the semi-local duality can be well satisfied by the $S$- and $P$-waves, 
mainly due to the cancellations of the $f_0(600)$ and $\rho(770)$ in the amplitudes. 
At large values of $N_C$, although the $f_0(600)$ resonance fades away in the complex energy plane and hence, barely 
contributes to the amplitudes~\cite{guo11prd,guo12plb}, another scalar strength survives and behaves like a standard $q\bar{q}$ resonance.  
The latter one corresponds to the singlet scalar $S_1$ which is part of the $f_0(980)$ for $N_C=3$.  
This is a crucial source to oppose the 
$\rho(770)$ contribution and hence guarantees the fulfillment of semi-local duality at large values of $N_C$.

\section*{Acknowledgements}

This work is partially funded by the grants MEC  FPA2010-17806 and the Fundaci\'on S\'eneca 11871/PI/09.
 We also thank the financial support from  the BMBF grant 06BN411, the EU-Research Infrastructure
Integrating Activity ``Study of Strongly Interacting Matter" (HadronPhysics2, grant No. 227431)
under the Seventh Framework Program of EU and the Consolider-Ingenio 2010 Programme CPAN (CSD2007-00042). 
Z.H.G. acknowledges CPAN postdoc contract in the Universidad de Murcia and 
financial support from the grants National Natural Science Foundation of China (NSFC) under contract No. 11105038, 
Natural Science Foundation of Hebei Province with contract No. A2011205093 and Doctor Foundation of Hebei Normal 
University with contract No. L2010B04.

\end{document}